\documentclass[%
 preprint,
 amsmath,amssymb,
 aps,
]{revtex4-1}

\usepackage{graphicx}
\usepackage{dcolumn}
\usepackage{bm}
\usepackage[flushleft]{threeparttable}
\usepackage{url}
\usepackage{amsmath}
\usepackage{amssymb}
\usepackage{braket}
\usepackage{multirow}
\usepackage{array}
\usepackage{booktabs}
\usepackage{float}
\usepackage{xcolor}
\usepackage{framed}
\usepackage{lipsum}
\usepackage{setspace} 

\colorlet{shadecolor}{yellow!25}
\newcolumntype{P}[1]{>{\centering\arraybackslash}p{#1}}

\begin{document}

\title{Theoretical Modeling of Tribochemical Reaction on Pt and Au Contacts: Mechanical Load and Catalysis}

\author{Yubo Qi, Jing Yang, and Andrew M. Rappe}
\affiliation{%
 The Makineni Theoretical Laboratories, Department of Chemistry,\\
 University of Pennsylvania, Philadelphia, PA 19104-6323 USA\\
}%
\date{\today}

\begin{abstract}
Micro--electro--mechanical system and nano--electro--mechanical system (MEMS and NEMS) transistors are considered promising for size--reducing and power--maximizing electronic devices. 
However, the tribopolymer which forms due to the mechanical load to the surface contacts affects the conductivity between the contacts dramatically.
This is one of the challenging problems that prevent widespread practical use of these otherwise promising devices. 
Here, we use density functional theory (DFT) to investigate the mechanisms of tribopolymer formation, including normal mechanical loading, the catalytic effect, as well as the electrochemical effect of the metal contacts. 
We select benzene select as the background gas, because it is one of the most common and severe hydrocarbon contaminants. 
Two adsorption cases are considered: one is benzene on the reactive metal surface, Pt(111), and the other is benzene on the noble metal, Au(111).
We demonstrate that the formation of tribopolymer is induced both by the mechanical load and by the catalytic effect of the contact.
First, benzene molecules are adsorbed on the Pt surfaces.
Then, due to the closure of the Pt contacts, stress is applied to the adsorbates, making the C--H bonds more fragile.
As the stress increases further, H atoms are pressed close to the Pt substrate and begin to bond with Pt atoms.
Thus Pt acts as a catalyst, accelerating the dehydrogenation process. When there is voltage applied across the contacts, the catalytic effect is enhanced by electrochemistry.
Finally, due to the loss of H atoms, C atoms become more reactive and link together or pile up to form tribopolymer.
By understanding these mechanisms, we provide guidance on design strategies for suppressing tribopolymer formation.

\end{abstract}

\pacs{Valid PACS appear here}
\maketitle
\section{Introduction}
Metal oxide semiconductor field effect transistors (MOSFETs) are the fundamental components in electronic logic devices. 
Moore's law predicts that the number of transistors in a defined size electronic device doubles every two years.
Effort toward shrinking the MOSFET scale is always ongoing, but a vital challenge is that the leakage current becomes unacceptably large as the MOSFET size decreases~\cite{Chen87p515}.
Besides, the minimum subthreshold swing $S= 60$ meV/decade sets a lower limit for energy dissipation in the MOSFET operation~\cite{taur1998fundamentals,Salahuddin08p405}.
To overcome these problems, microelectromechanical system (MEMS) switches are being developed~\cite{Loh12p283,Spencer11p308,Lee13P36,Sinha12p484,Czaplewski09p085003}. 
The source and drain electrodes are separated by an air gap, and they can also be connected by an adjustable conductive bar.
Such an electronic device has nearly zero leakage current, and the minimum subthreshold swing also breaks the $S= 60$ meV/decade limitation. 
However, a prime concern about this approach is the stability of the electrical contact resistance (ECR);
a hydrocarbon contaminant forms on the surfaces of the electrodes and on the conductive bar after many switching cycles. 
This tribopolymer may increase the ECR beyond the tolerance of logic applications~\cite{Dickrell07p75,Jensen05p935,Brand13p341,Brand13p1248,Hermance58p739}.

The study of the reliability of MEMS devices has been a fast--moving field~\cite{VanSpengen03p1049}.
In MEMS devices, Pt and Au are two common electrodes in use. 
The benefits of using Au electrodes are corrosion retardation, low resistivity, and ease of deposition~\cite{Hyman99p357,Patton05p215,Yang09p287}. 
However, Au is a soft metal with low cohesive energy,
so surface wear and material transfer are the main drawbacks of Au contacts. 
Also, some contamination can still form after electrical cycling~\cite{Dickrell07p75,Tringe03p4661}. 
As for Pt contacts, previous experiments showed that tribopolymer is more prone to form on Pt than on Au.
Wabiszewski ${et al}$. used atomic force microscopy (AFM) to mimic the
Pt/Pt electrical contacts. The ECR after cycling is six orders of magnitude higher than before ~\cite{Wabiszewski13p1}. 
Brand $et\ al.$ observed considerable amounts of tribopolymer formation on Pt contacts, leading to a dramatic increase of ECR~\cite{Brand13p341,Brand13p1248}.

For Pt contacts, the rate of formation and the amount of tribopolymer are related to how the MEMS switches are operated~\cite{Brand13p341,Brand13p1248}. 
Three types of switching-cycle mechanisms are considered in their experimental setup: 
cold switching, hot switching, and mechanical switching. In cold switching, voltage is applied after the two contacts are closed, and removed before the contacts open. 
In hot switching, voltage is applied across the device the entire time, whether the contacts are closed or open. 
In mechanical switching, in order to better understand the role of electrochemistry relative to other effects, the contacts undergo cycles of closing and opening without applying any voltage.  
When a voltage of 5 V was applied across the contacts during the cold--switching cycle, a large amount of tribopolymer was generated.
However, interestingly, even in mechanical switching, the tribopolymer is still formed, but less compared to the cold switching protocol. 
In this work, we applied a computational compression experiment to study the effects of mechanical stress and catalysis on tribopolymer formation.

\section{Methodology}
It is believed that the tribopolymer results from the polymerization of hydrocarbon gases from the atmosphere or from packaging. 
Among them, benzene is the one which causes the most severe contamination~\cite{Hermance58p739}.
Therefore, benzene is selected as the background gas in our study.
In order to investigate tribopolymer formation on Pt contacts in a benzene atmosphere, 
we conducted compression computational experiments with first--principles density functional theory (DFT) calculations. 
Au contacts are also studied as a control system to illustrate the tribopolymer formation mechanism.
All calculations are performed with the plane--wave DFT package QUANTUM--ESPRESSO~\cite{Giannozzi09p395502etal} using the generalized gradient approximation (GGA) exchange correlation functional. 
A plane--wave cutoff energy of $E_{\text{cut}}\approx680$ eV and an 8$\times$8$\times1$ Monkhorst--Pack $k$--point mesh~\cite{Monkhorst76p5188} are used for energy and force calculations. 
All atoms are represented by norm-conserving pseudopotentials generated by the OPIUM~\cite{Opium} code with a plane-wave cutoff, $\hbar^2q_{\text{c}}^2/2m_e\approx680$ eV.

\begin{figure}[htbp]
\includegraphics[width=16.0cm]{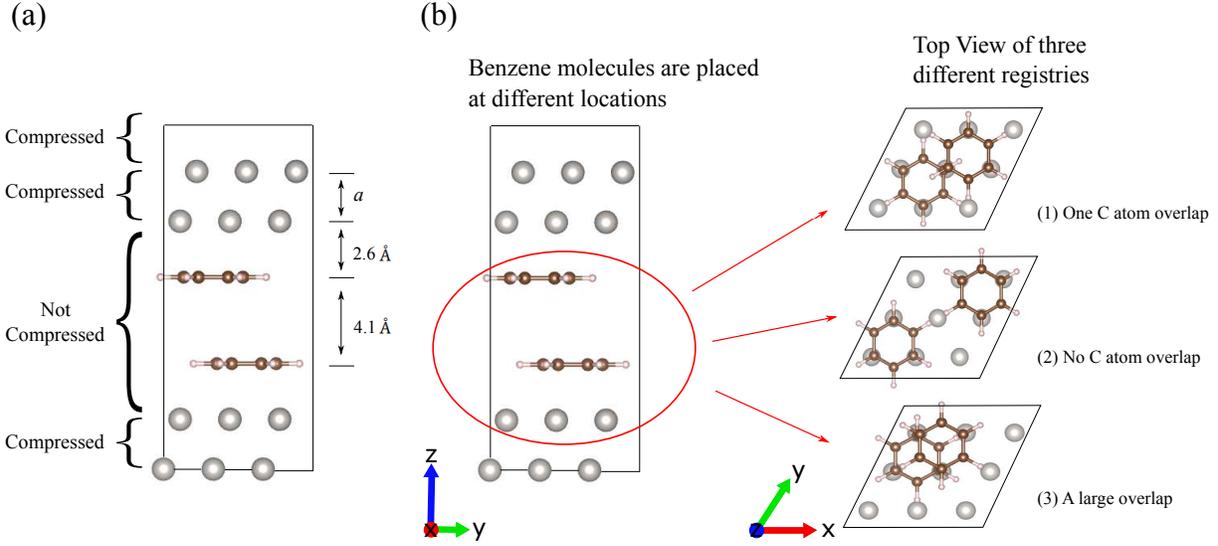}
\caption{Ball--and--stick model of the initial registry of two benzene molecules. 
(a) Side view of the compression supercell setup with 4--layer Pt (111) slab and two benzene molecules. 
Compression is initially applied to the metal interlayer spacings, leaving the center vacuum uncompressed (Pt = light grey, C = brown, H = light pink).  
(b) Top view of three different benzene initial registries on upper and lower Pt slabs. The left panel shows the entire supercell.
On the right panel, all benzene molecules sit on the hollow site (three C atoms on top of Pt atoms and three in the Pt 3--fold hollow) of the Pt surface with different horizontal intermolecular distances:
(1) two carbon directly overlapped, (2) the furthest, no overlap at all, (3) significantly overlapped.}
\label{f12}
\end{figure}

A supercell geometry is used with periodically separated metal slabs and two benzene molecules per surface cell between the slabs to model metal contacts in a benzene atmosphere.
A four--layer metal slab gives converged surface calculations. 
FIG.~\ref{f12} shows the initial setup of the supercell structures. 
The vertical distance between the two parallel benzene molecules is selected to be 4.1~\AA, so that the two benzene molecules have little interaction. 
In the initial structure, the benzene molecules are intact without any deformation.
The distance between one benzene molecule and its closest metal contact is 2.6~\AA. 
At this distance, the benzene molecule has an attractive adsorption force toward the contact, and no stress is required to press the molecule to such a position.  
After the relaxation, the two benzene molecules are strongly chemisorbed on the top and bottom contacts. 
Once the adsorption system is fully optimized, the value of the distance between each pair of neighboring metal atomic layers, $a$, is decreased by $\Delta{a}$.
Since there are three intervals in the four--layer slabs,
the height of the supercell is reduced by $3\Delta{a}$ in total in each compression step.
At the beginning, the value of $\Delta{a}$ is selected as 0.2 \AA.
Once any significant deformation of the adsorbates is observed, we change $\Delta{a}$ to 0.02 \AA,
in order to have a subtle observation of the structural changes correlated with the height of the supercell.
Then, we optimize the compressed system, and after the relaxation we repeat this compression and optimization process. 
Our manual adjustment of the supercell is analogous to applying stress to the adsorbed benzene.
As the compression process goes on, the normal load (stress) increases.
The average stress is calculated from the energy of the system and the dimension of the supercell
\begin{equation}
\sigma=-\frac{1}{S}\frac{\partial{E}}{\partial{z}}
\end{equation}
where $\sigma$ is the stress along the $z$ direction of the system, $E$ is the energy of the system, and $S$ is the area of the surface. 
Since we have optimized the structure to equilibrium, 
the stress in the supercell is approximately uniform and the stress on the molecules is the same as the average one.

\section{Results}

In Pt--benzene systems, we choose three different registries as shown in FIG. \ref{f12}. 
In registry (1), there is one carbon atom overlapping between the two benzene molecules from the top view. 
In the second registry, there is no carbon atom overlap.
In registry (3), the two benzene molecules have a relatively large overlap.

\begin{figure}[htbp]
\includegraphics[width=15.0cm]{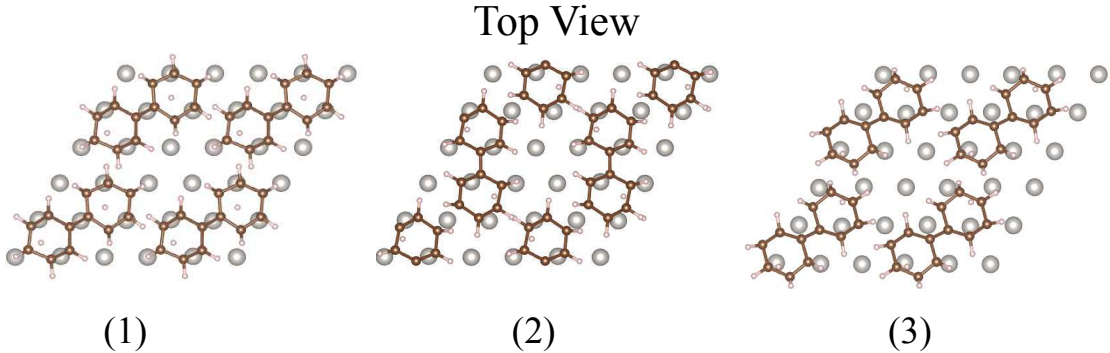}
\caption{Top views of the biphenyl--like structure formed after the height of the supercell is reduced to 11.06~\AA\ for the three different registries portrayed in Fig. \ref{f12}.
For clarity, only the adsorbates and the first layer of Pt atoms beneath are shown.}
\label{f3}
\end{figure}

During the relaxation, the electrodes will recover the lattice structure to a new equilibrium condition,
and hence will expand and squeeze the center vacuum space. The adsorbed molecules at the center will sense the stress applied by the two closing contacts. 
We conduct compression computational experiments and observe that for all three different registries, when the height of the supercell is reduced to 11.06 \AA, 
chemical reaction occurs: some hydrogen atoms dissociate from the adsorbed benzene molecules, making them less saturated.
The carbon atoms which lose their bonded hydrogen atoms form a C--C bond making a biphenyl--like structure, as shown in FIG. \ref{f3}. 

\begin{figure}[htbp]
\includegraphics[width=12.0cm]{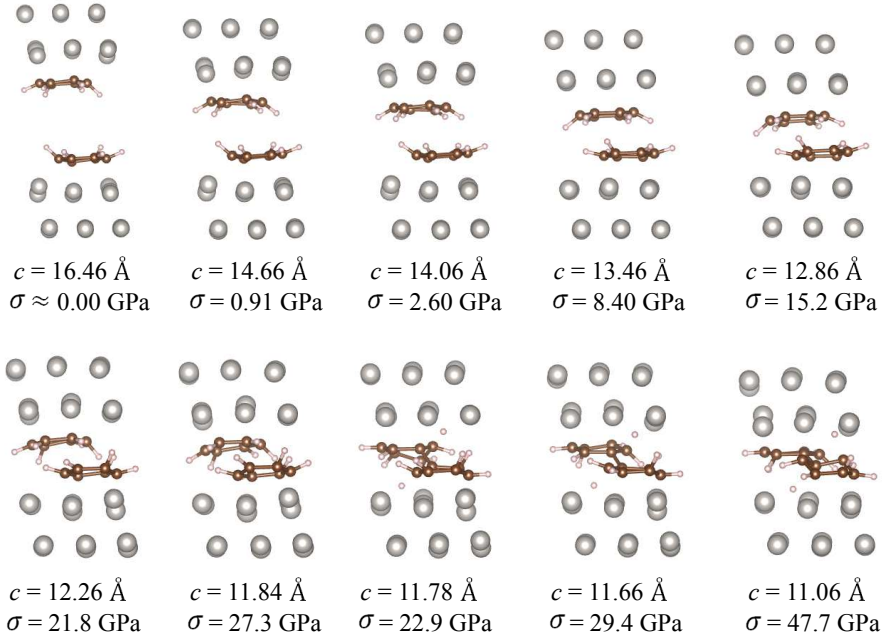}
\caption{Ball--and--stick model of the entire compression process of the supercell with benzene on the Pt(111) surface. 
$c$ is the length of the supercell along the $z$--direction. $\sigma$ is the stress of the supercell at each $c$ value. 
At $c=11.78$ \AA\ and $\sigma=22.9$ GPa, the biphenyl--like structure forms.}
\label{f4}
\end{figure}

In the following part, we take a more detailed look at the formation of the biphenyl--like structure.
For the structure without any compression, the benzene molecules adsorb on the surface and distort to become non--planar.
The adsorption is at the hollow site, as it is the most favorable position~\cite{Saeys02p7489}, as shown in FIG.~\ref{f4} (a).
The C--C bonds are around 1.5~\AA, which is approximately the length of C--C single bond.
These indicate the chemisorption of benzene on Pt(111) surface. 
The bonding between C and Pt induces the H atoms to move away from the surface, indicating $sp^3$ C character.
The C--Pt bond distances are approximately~2.1~\AA.
We should also note that before the compression (no external stress applied), the distance between one H atom and its closest Pt atom is approximately 2.7~\AA, 
which is much larger than the length of a Pt--H bond (1.7 \AA)~\cite{Walter04p265}.
There is no chemical bond between Pt and H atoms. 

\begin{figure}[htbp]
\includegraphics[width=15.0cm]{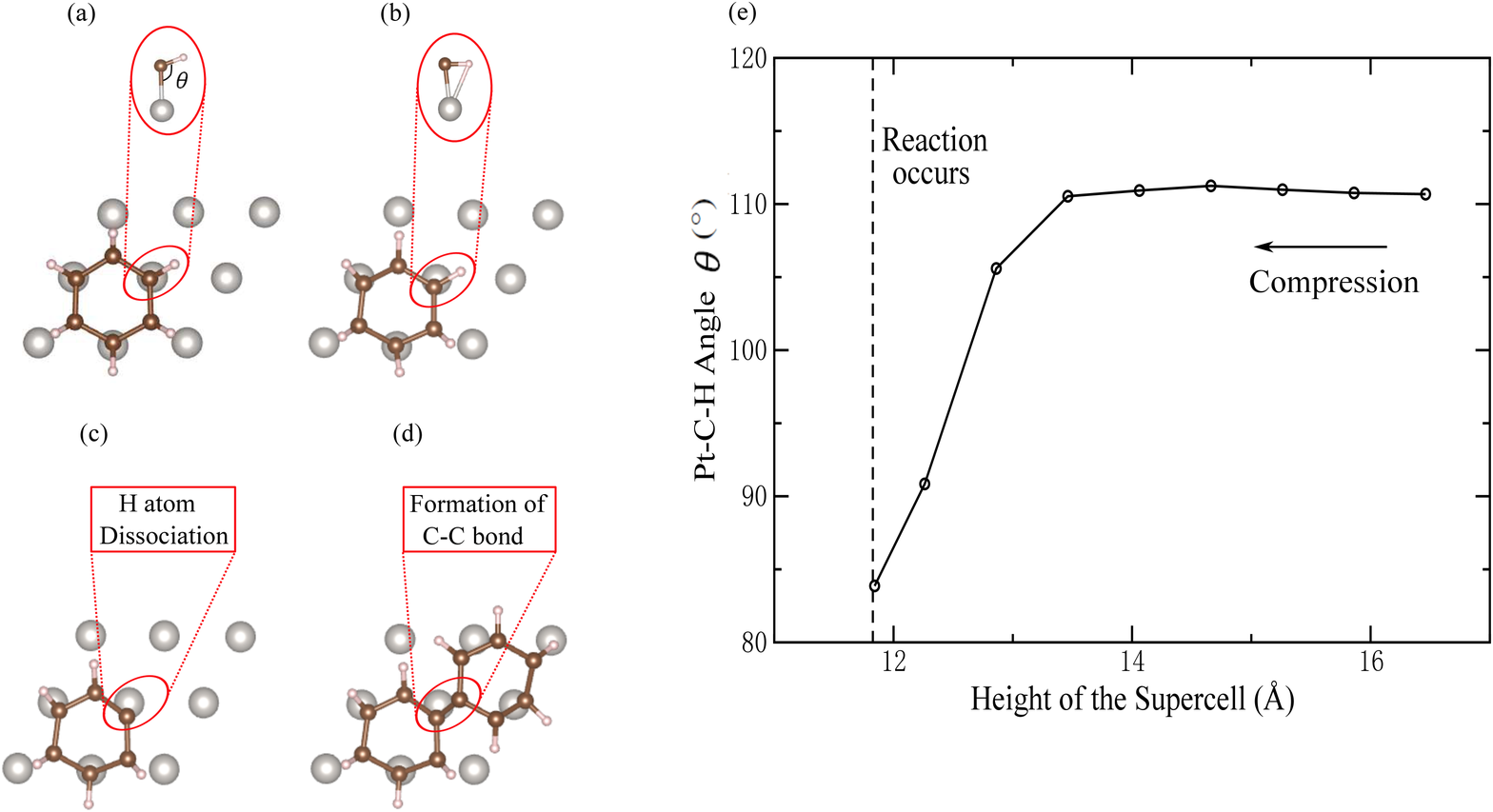}
\caption{Ball--and--stick model of detailed benzene bonding geometry and reactivity on Pt (111). 
(a) Benzene molecule adsorbed at the hollow site. The lengths of the Pt--C bonds are 2.1~\AA, and the C--C bonds are 1.54~\AA; 
(b) Part of the structure of the Pt--benzene system with $c=11.84$ \AA\ and $\sigma=27.3$ GPa. 
One H atom is pressed close to a Pt atom;
(c) The H atom dissociates from the adsorbed benzene;
(d) A biphenyl--like structure forms;
(e) The plot of supercell height $vs.$ the Pt--C--H angle. 
As the compression goes on, the angle is reduced, indicating that the H atoms are pressed toward the Pt contact.}
\label{f8}
\end{figure}

Taking the first registry as an example, the changes of the structure with the height of the supercell $c$ and the resulting $\sigma$ are shown in FIG.~\ref{f4}. 
When the stress increases to 27.3 GPa, it can been seen that one H atom is pushed close to a Pt atom, as shown in FIG.~\ref{f8}. 

As we compress the system further, a C--H bond breaks, and the H atom is attached to the Pt substrate.
The adsorption position of the dissociated H atom is the hollow site, which is the most stable one~\cite{Watson01p4889}.
Two C atoms which lose their H atoms form a C--C bond, and connecting the carbon rings builds a biphenyl--like structure.
The adsorbed benzene molecules are on the way to more dramatic change: 
(1) H atoms can fall off due to further compression. The detached atoms may form H$_2$ and escape into the atmosphere, 
which indicates that the dehydrogenation is a non--reversible process; 
(2) C atoms which lose their H atoms become less saturated and more reactive.
As these processes continue, more hydrogen atoms may fall off, and reactive C atoms link with others or even pile up, 
forming a two dimensional or three dimensional C--rich network. 
The biphenyl--like linked rings (Fig. 4) can be viewed as the primary step of tribopolymer formation.

To further investigate the dehydrogenation and polymerization processes,
we use the nudged elastic band (NEB)~\cite{Henkelman00p9901} method to calculate the energy barrier at different unit cell heights, as shown in Fig. 5. 
Initial structures with different supercell heights provide different stress/mechanical loads. 
The calculated unit cell height dependence of reaction energies is shown in FIG.~\ref{f5_2}.
For stress below 26.6 GPa, there is a linear relation between stress and reaction barrier, which is consistent with a Bell--Evans--Polanyi (BEP) relationship~\cite{Black11p1655,Bell78p618,Konda11p164103} as expressed  
\begin{equation}
\Delta\left(\Delta{E}_{\rm{act}}\right)=F\cdot{\Delta}d
\end{equation}
where $\Delta\left(\Delta{E}_{\rm{act}}\right)$ is the change of activation energy $\Delta{E_{\rm{act}}}$, $F$ is the applied force, 
which is proportional to the stress, and $\Delta{d}$ is the distance that the force $F$ moves though.
As we increase the stress and the dimension of the system shrinks by $\Delta{d}$, 
a work
\begin{equation}
W=F\cdot{\Delta}d
\end{equation}
is applied to the system.
This energy is added to the adsorbates, promoting the reaction, as shown in Fig. 5 (b). 
If the stress increases further, the energy barrier drops from 1.2 eV to zero within just a 2 GPa period.
This breakdown of the BEP relationship is attributed to the catalysis effect of Pt.
Compression not only adds energy to the adsorbates, 
but also presses hydrogen atoms close to the Pt substrate, changing the reaction pathway.
Pt, as a catalyst, can weaken the C--H bond and reduce the reaction energy barrier.
The stronger the Pt--H interaction is, the lower the energy barrier is.

The reaction occurs when the stress is 27.3 GPa. 
In the following part, we demonstrate that the threshold stress can be reduced by including the electrochemical effect.
Above 26.6 GPa, the relationship of the supercell height and energy barrier deviates from linearity.
We deduce that from this point, Pt begins to affect the dehydrogenation as a catalyst.
The chemical reaction can be expressed as
\begin{equation}
\rm{C}_6\rm{H}_6\ \rightarrow\ \rm{C}_{6}\rm{H}_{5}\bm{\cdot}\ +\ \left(\rm{H}^{+}+e^{-}\right).
\end{equation}
And the reaction energy change due to applied potential $U$ is~\cite{Martirez15p2939}
\begin{equation}
\Delta{E}_{\rm{act}}^{*}=\Delta{E}_{\rm{act}}-n\cdot{e}\cdot{U}
\end{equation}
where $e$ is the electron charge and $n$ is the number of electrons involved in the reaction.
At 26.6 GPa, the energy barrier is 1.4 eV, and in experiments, the applied voltage (5.0 V) during device switching is larger than 1.4 V.
Therefore, due to the electrochemical effect, the H atom can get one electron and leave the benzene as soon as it makes electrical contact with the electrode.

The process of adsorbed benzene molecules losing H atoms and linking together by C bonds can be summarized as:
(1) benzene molecules are adsorbed on the Pt surfaces;
(2) due to the closure of the Pt contacts, stress is applied to the adsorbate, flattening it and making C--H bonds easier to break.
We emphasize that while the H atoms are not close to the Pt substrate, 
the fragility of adsorbed benzene mainly results from mechanical load (rather than catalytic effect), and the activation energy--stress relationship obeys the BEP relationship;
(3) As the stress increases further, the benzene adsorbates become flat enough and the H atoms begin to be attracted to the Pt atoms.
Pt weakens the C--H bond and accelerates the dehydrogenation process, as shown in Fig. 5 (b). This effect can also be enhanced by electrochemistry.

\begin{figure}[htbp]
\includegraphics[width=17.0cm]{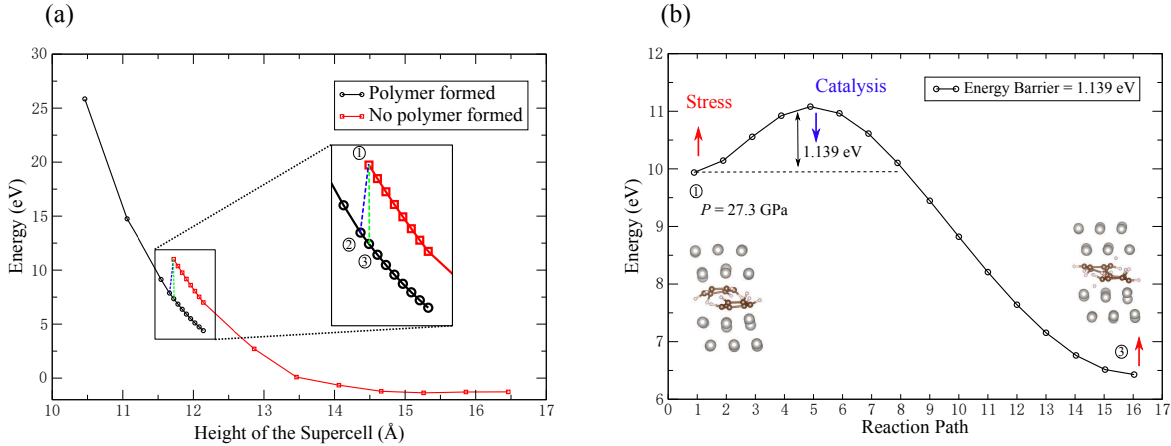}
\caption{Energetic behavior of the compression process. The energies are relative energies, with the sum of the energies of a clean Pt slab and two free benzene molecules as zero.
(a) Total energy ${vs.}$ height of the supercell during the compression process of Fig.~\ref{f4}. 
Compression before the polymer formation is shown in red, whereas the continued compression after the formation of the polymer and expansion of the supercell shows in black. 
The subplot shows the detailed stress effect on the polymer formation process. \textcircled{1} is the energy of endpoint of the supercell structure one step before the formation of polymer.
The energy drop to \textcircled{2} indicates the polymer formation. 
Continued compression of the cell does not lead to additional new species formation. \textcircled{3} is an expansion step back up to the same supercell volume as \textcircled{1}. 
(b) The energy barrier to form the biphenyl tribopolymer as in Fig. \ref{f4}, calculated by NEB method. 
The supercell configurations at \textcircled{1} and \textcircled{3} are taken to be the initial and final structures of the reaction.
Schematic model demonstrating the change of the energy barrier is also shown.
At the beginning, mechanical load raises the energy levels of the initial and final states (red arrows). The increased energy is proportional to the stress.
Then, the height of the barrier drops (blue arrow), due to the catalytic interaction of Pt and H.}
\label{f5}
\end{figure}

\begin{figure}[htbp]
\includegraphics[width=9.0cm]{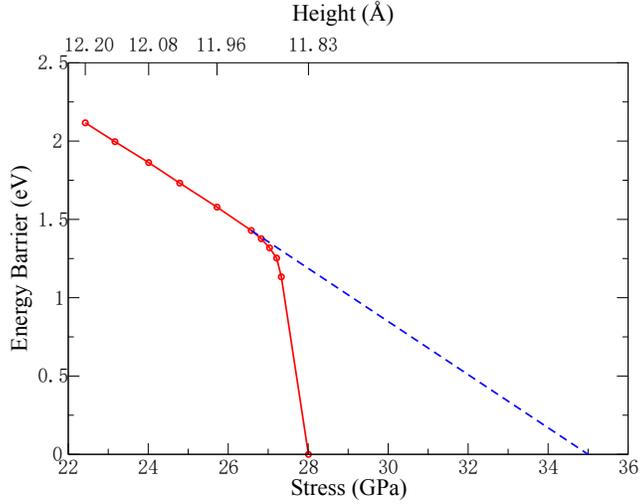}
\caption{Energy barriers for benzene on Pt at different stresses. 
Above 26.6 GPa, the relationship of energy barrier and stress is no longer linear, demonstrating that the catalysis effect begins to act.
The dashed blue line is extrapolated from the linear part. This line and the $x$--axis intersect at 35 GPa, 
indicating the estimated polymerization stress without catalysis.}\label{f5_2}
\end{figure}


In the following section, computational compression experiments of a Au--benzene system are analyzed as a contrast with Pt,  
highlighting the importance of adsorption and catalysis in polymerization.
\begin{figure}[htbp]
\includegraphics[width=15.0cm]{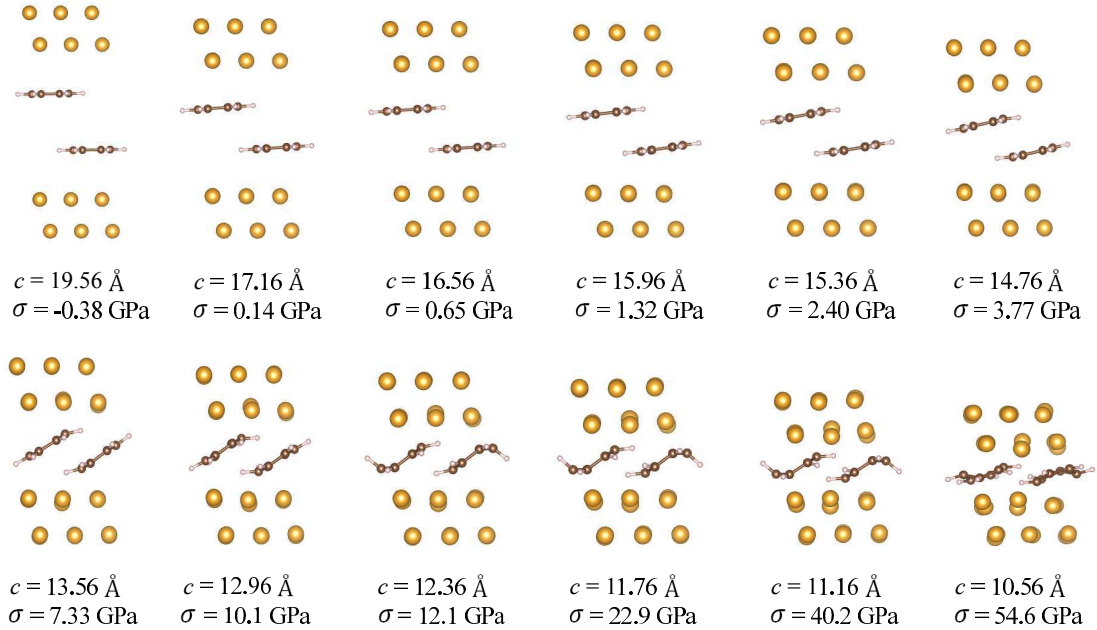}
\caption{Ball--and--stick model of the compression of benzene on the Au(111) surface.
$c$ is the supercell height along the $z$--axis, and $\sigma$ is the stress of the supercell. 
Different from the benzene on Pt(111) case, no polymer formation is seen during the compression.}
\label{f6}
\end{figure}
In FIG. \ref{f6}, the changes of the Au--benzene structure during the compression are shown.
At the outset, benzene molecules are not strongly adsorbed on the contact surfaces, with a -0.1 eV adsorption energy, 
much weaker than on Pt (-1.9 eV). 
After normal load is applied, benzene molecules spontaneously glide and avoid each other, due to the lack of strong adsorption or fixed location.
As the compression increases, the benzene molecules begin to distort, because of the reduction of free space and applied stress.
Different from the benzene on Pt contacts case, there is no C--H bonds broken.
Even though the benzene ring is broken at the high stress cases and Au--C bonds are able to form,
no polymerization is observed, even when the height of the supercell is reduced to 10.56 \AA\ with stress of 54.6 GPa,
which is far beyond the requirement of benzene polymerization on Pt contacts ($\approx$ 27 GPa).

As we know, Au is the noblest metal, possessing a completely filled $d$--band, and the $d$--band center is far from the Fermi level~\cite{Hammer95p238}.
As a result, the bonds that form between Au and adsorbates have anti--bonding orbitals that are partially or mostly filled, 
indicating that the bonding itself is not stable.
Therefore, Au contacts cannot adsorb molecules tightly, so there is nothing to stop intermolecular gliding, and Au does not catalyze the dehydrogenation reaction.

If the concentration of benzene is low, 
the molecules tend to escape from the center space between the two Au contacts,
making tribopolymer formation less likely.
For the high concentration case,
if lots of benzene molecules are trapped in a certain region during the switching,
the mechanical load can eventually induce the polymerization.
Previous theoretical study demonstrates that benzene molecules can transform to a polymeric phase due solely to pressure~\cite{Wen11p9023}, 
but the required polymerization pressure is relatively high, above 80 GPa.

The main differences between the Pt and Au cases are the adsorption and dehydrogenation.
Adsorption is important, because it not only prevents the repulsion and avoidance of molecules, but also breaks the strong and stable delocalized $\pi$ orbital.
For the Au contacts with little adsorption, the delocalized $\pi$ system is crushed only by the mechanical load, 
and there is a stricter requirement for this process, such as a higher molecule concentration or larger mechanical load.
However, for the polymerization on Pt case, the strong delocalized $\pi$ system is broken by the adsorption onto the substrate.
Polymerization is induced by the breaking of the C--H bonds, and this process is much easier.

For Pt contacts, computational compression experiments demonstrate that polymerization can occur under stress of 26.6 GPa.
Compared with the pure benzene polymerization stress 80 GPa, it is much closer to the experimental situation, for which the stress between contacts is estimated at several GPa.
Besides, only uniaxial stress is investigated in this study.
Shear may also assist the dehydrogenation, accelerate the polymerization, and hence lower the polymerization threshold stress.

\section{Conclusion}

In this study, the mechanism of tribopolymer formation has been studied by computational compression experiments.
Benzene molecules adsorb on the surfaces of contacts, and then original chemical bonds can be broken due to mechanical load. 
Dehydrogenation makes adsorbed benzene molecules less saturated, so they bond with others to form tribopolymer. 
In a certain stress range, the reaction energy decreases with mechanical load, and the linear relationship between $E_{\rm{act}}$ and $P$ follows the BEP relationship.
When the stress is large enough to press H atoms close to the Pt substrate, a catalytic effect takes over and the energy barrier decreases to zero more rapidly as stress increases.
Our study provides a detailed analysis of the process and mechanism of the initial stage of tribopolymer formation on contacts in MEMS and NEMS.

\section*{ACKNOWLEDGMENTS}
Y.Q. was supported by the U.S. National Science Foundation, under grant DMR1124696.
J.Y. was supported by the U.S. National Science Foundation, under grant CMMI1334241.
A.M.R. was supported by the U.S. Department of Energy, under grant DE--FG02--07ER15920.
Computational support was provided by the National Energy Research Scientific Computing Center of the Department of Energy.

\bibliography{rappecites}

\end{document}